\documentclass[preprint,showpacs,prl]{revtex4}
\usepackage{epsf}
\usepackage{dcolumn}
\usepackage{bm}
\everymath{\displaystyle}
\begin{document}

\title{Quantum-mechanical description of spin-1 particles with
electric dipole moments}

\author{Alexander J. Silenko}

\affiliation{Research Institute for Nuclear Problems, Belarusian
State University, 220030 Minsk, Belarus}

\date{\today}

\begin {abstract}
The Proca-Corben-Schwinger equations for a spin-1 particle with an
anomalous magnetic moment are added by a term describing an
electric dipole moment, then they are reduced to a Hamiltonian
form, and finally they are brought to the Foldy-Wouthuysen
representation. Relativistic equations of motion are derived. The
needed agreement between quantum-mechanical and classical
relativistic equations of motion is proved. The scalar and tensor
electric and magnetic polarizabilities of pointlike spin-1
particles (W bosons) are calculated for the first time.
\end{abstract}

 \maketitle

\section{Introduction}

The discovery of electric dipole moments (EDMs) is one of the main
goals of contemporary physics. Such a discovery would go beyond
the Standard Model and open a window to new physics. To disclose
the EDM, one needs to find an anomaly in spin dynamics. The
classical spin dynamics of a particle with an anomalous magnetic
moment (AMM) and an EDM was determined many years ago
\cite{Nelson:1959zz}, but the corresponding quantum-mechanical
equations have been derived only for spin-1/2 particles
\cite{NPR,RPJ}.

One of the experimental priorities is a search for the EDM of the
deuteron \cite{Semertzidis:2003iq,dEDM,Lehrach:2012eg} whose spin
is 1. While the classical and quantum theories of spin motion
should agree, a proper quantum-mechanical consideration of spin-1
particles/nuclei is also necessary. We perform such a
consideration based on the Proca equations \cite{Pr} with an
additional term included by Corben and Schwinger \cite{CS}. We
generalize the above Proca-Corben-Schwinger (PCS) equations to
take also into account the EDM, then bring the generalized
equations to a Hamiltonian form and perform the relativistic
Foldy-Wouthuysen (FW) transformation. Unlike the original FW
approach \cite{FW}, we use the method \cite{JMP,PRA} that enables
transition to the FW representation for relativistic particles in
external fields. This allows us to find a relativistic operator
equation of spin motion and easily determine its classical limit.
This result provides a deficient quantum-mechanical basis for the
deuteron EDM experiment.

Finally, we calculate for the first time scalar and tensor
electric and magnetic polarizabilities of pointlike
(structureless) spin-1 particles.

We use the system of units $\hbar=1,~c=1$.

\section{Basic equations}

Proca equations \cite{Pr} for spin-1 particles with the
Corben-Schwinger term \cite{CS} have the form
\begin{equation} U_{\mu\nu}=D_\mu U_\nu-D_\nu U_\mu, ~~~ \mu,\nu=0,1,2,3,
\label{eqCorSc}
\end{equation}
\begin{equation} D^\mu U_{\mu\nu}-m^2 U_\nu+ie\kappa U^\mu F_{\mu\nu}=0,
\label{eqCorSh}
\end{equation} where $D_\mu=\partial_\mu+ieA_\mu$ is the covariant derivative, $A_\mu$ is the
four-potential, $F_{\mu\nu}$ is the electromagnetic field tensor,
and $ U_{\mu\nu}=-U_{\nu\mu}$. The Corben-Schwinger term is
proportional to $\kappa=g-1$, where $g=2m\mu/(es)=2m\mu/e$ for
spin-1 particles. Since the Proca equations correspond to $g=1$,
this term describes not only the AMM but also a part of the normal
($g=2$) magnetic moment \cite{YB}. Spin-1 particles can be also
described by the Duffin-Kemmer-Petiaux equation \cite{DKP},
Stuckelberg equation \cite{St}, multispinor Bargmann-Wigner
equations \cite{BW}, and other equations.

Since the spin of Proca particles has three components, six
components of the wave function are independent. Spatial
components of Eq. (\ref{eqCorSc}) and a time component of Eq.
(\ref{eqCorSh}) can be expressed in terms of the others. As a
result, the equations for the ten-component wave function can be
reduced to the equation for the six-component one (Sakata-Taketani
transformation \cite{SaTa}). The distinctive feature of this
transformation is that it obtains expressions for $U_0$ and
$U_{ij}~(i,j=1,2,3)$ which do not contain the time derivative and
then it substitutes them into equations for the remaining
components. From Eq. (\ref{eqCorSh}) we have
$$ U_0=\frac{1}{m^2}\left( D^i U_{i0}+ie\kappa U^i F_{i0}\right).$$
Next we introduce two vector functions, $\bm\phi$ and $\bm U$,
whose components are given by $iU_{i0}/m$ and $U^i$ and form the
six-component Sakata-Taketani wave function
$$\Psi =\frac{1}{\sqrt2}\left(\begin{array}{c} \bm \phi+\bm U \\
\bm \phi-\bm U \end{array}\right).$$

As the generalized Sakata-Taketani equation can be expressed in terms of
spin-1 matrices \cite{YB}, the wave function of this equation
is similar to a Dirac
bispinor. The general form of the Hamiltonian in the
Sakata-Taketani representation obtained by Young and Bludman
\cite{YB} is given by
\begin{equation}\begin{array}{c}
{\cal H}=e\Phi+\rho_3 m+i\rho_2\frac{1}{m}(\bm S\cdot\bm
D)^2\\-(\rho_3+i\rho_2) \frac{1}{2m}(\bm D^2+e\bm S\cdot\bm B)-
(\rho_3-i\rho_2) \frac{e\kappa}{2m}(\bm S\cdot\bm B)\\-
\frac{e\kappa}{2m^2}(1+\rho_1)\biggl[(\bm S\cdot\bm E)(\bm
S\cdot\bm D)-i \bm S\cdot[\bm E\times\bm D]-\bm E\!\cdot\!\bm
D\biggr]\\ +\frac{e\kappa}{2m^2}(1-\rho_1)\biggl[(\bm S\cdot\bm
D)(\bm S\cdot\bm E)-i \bm S\cdot[\bm D\times\bm E]-\bm
D\!\cdot\!\bm
E\biggr]\\
-\frac{e^2\kappa^2}{2m^3}(\rho_3-i\rho_2)\biggl[(\bm S\cdot\bm
E)^2- \bm E^2\biggr],
\end{array} \label{eq15} \end{equation}
where $\bm S$ is the $3\times3$ spin matrix, $\rho_i~(i=1,2,3)$
are the $2\times2$ Pauli matrices, $\kappa=const$, $\bm E$ is the
electric field strength, and $\bm B$ is the magnetic field
induction. We do not consider a nonintrinsic quadrupole moment
included in Ref. \cite{YB}. Denotation $\rho_iS_j$ means the
direct product of two matrices. For spin-1 particles, the
polarization operator is equal to $\bm\Pi=\rho_3\bm S$. It is
analogous to the corresponding Dirac operator which can be written
in a similar form (see Ref. \cite{ST}): $\bm\Pi=\rho_3\bm\sigma$.

In Refs. \cite{JETP,arXiv}, Hamiltonian (\ref{eq15}) has been
transformed to the FW representation for relativistic particles in
electric and magnetic fields with allowance for derivatives of the
electric field strength. The terms proportional to the derivatives
of the magnetic field induction have not been calculated.

\section{Inclusion of electric dipole moments}

To describe the EDMs of spin-1/2 particles, the terms proportional
to the $\gamma^5$ matrix can be added to the Lagrangian and the
Dirac equation \cite{Feng:2001sq}. It has been shown in Ref.
\cite{RPJ} that there exists another way to include
the EDMs with the tensor $G_{\mu\nu}=(\bm B,-\bm E)$ 
dual to the electromagnetic field one, $F_{\mu\nu}=(\bm E,\bm B)$.
In this case, the Lagrangians describing the AMM and EDM become
very similar and are given by \cite{RPJ}
\begin{equation} {\cal L}_{AMM}=\frac{\mu'}{2}\sigma^{\mu\nu}F_{\mu\nu},
~~~~~ {\cal L}_{EDM}=-\frac{d}{2}\sigma^{\mu\nu}G_{\mu\nu},
\label{eq5} \end{equation} where $d$ is the EDM of the particle.
The generalized Dirac-Pauli equation assumes the form \cite{RPJ}
\begin{equation} \biggl[\gamma^\mu\pi_\mu-m+\frac{\mu'}{2}\sigma^{\mu\nu}F_{\mu\nu}
-\frac{d}{2}\sigma^{\mu\nu}G_{\mu\nu}\biggr] \Psi=0.\label{gDPeq}
\end{equation}
The terms describing the contributions of the AMM and EDM to the
Hamiltonian are transformed into each other using the
substitutions $\bm B\rightarrow \bm E,~ \bm E\rightarrow-\bm B,~
\mu'\rightarrow d$. The corresponding relativistic FW Hamiltonian
and equations of motion have been derived in Ref. \cite{RPJ} by
the method developed in Ref. \cite{JMP}.

Similarly, we can supplement the Lagrangian of spin-1 particles
\cite{YB} containing the AMM term, ${\cal
L}_{AMM}=(ie\kappa/2)(U_\mu^\dag U_\nu-U_\nu^\dag
U_\mu)F^{\mu\nu}$, with the EDM one:
\begin{equation} {\cal
L}_{EDM}=-(ie\eta/2)(U_\mu^\dag U_\nu-U_\nu^\dag
U_\mu)G^{\mu\nu},\label{LEDM}
\end{equation}
where $\eta=2dm/(es)=2dm/e$. The corresponding generalized PCS
equations read
\begin{equation}\begin{array}{c} U_{\mu\nu}=D_\mu U_\nu-D_\nu U_\mu,
\\
D^\mu U_{\mu\nu}-m^2 U_\nu+ie\kappa U^\mu F_{\mu\nu}-ie\eta U^\mu
G_{\mu\nu}=0.
\end{array} \label{gPCS}
\end{equation}

This generalization brings Eq. (\ref{eq15}) to the form
\begin{equation}\begin{array}{c}
{\cal H}=e\Phi+\rho_3 m+i\rho_2\frac{1}{m}(\bm S\cdot\bm
D)^2\\-(\rho_3+i\rho_2) \frac{1}{2m}(\bm D^2+e\bm S\cdot\bm B)-
(\rho_3-i\rho_2) \frac{e\kappa}{2m}(\bm S\cdot\bm B)\\-
\frac{e\kappa}{2m^2}(1+\rho_1)\biggl[(\bm S\cdot\bm E)(\bm
S\cdot\bm D)-i \bm S\cdot[\bm E\times\bm D]-\bm E\!\cdot\!\bm
D\biggr]\\ +\frac{e\kappa}{2m^2}(1-\rho_1)\biggl[(\bm S\cdot\bm
D)(\bm S\cdot\bm E)-i \bm S\cdot[\bm D\times\bm E]-\bm
D\!\cdot\!\bm
E\biggr]\\
-\frac{e^2\kappa^2}{2m^2}(\rho_3\!-\!i\rho_2)\biggl[(\bm
S\!\cdot\!\bm E)^2- \bm E^2\biggr] - (\rho_3\!-\!i\rho_2)
\frac{e\eta}{2m}(\bm S\!\cdot\!\bm E)\\+
\frac{e\eta}{2m^2}(1+\rho_1)\biggl[(\bm S\cdot\bm B)(\bm S\cdot\bm
D)-i \bm S\cdot[\bm B\times\bm D]-\bm B\!\cdot\!\bm D\biggr]\\
-\frac{e\eta}{2m^2}(1\!-\!\rho_1)\biggl[(\bm S\!\cdot\!\bm D)(\bm
S\!\cdot\!\bm B)-i \bm S\cdot[\bm D\!\times\!\bm B]-\bm
D\!\cdot\!\bm B\biggr],
\end{array} \label{eqEDM} \end{equation}
where only first-order terms in $\eta$ are taken into account.

The use of the appropriate method of the FW transformation for
relativistic particles in external fields \cite{JMP,PRA} leads to
the following FW Hamiltonian:
\begin{equation} \begin{array}{c}
{\cal
H}_{FW}=\rho_3\epsilon'+e\Phi+\frac{e}{4m}\left[\biggl\{\biggl(\frac{g-2}{2}\right.\\+
\frac{m}{\epsilon'+m}\biggr)\frac{1}{\epsilon'}, \bigl(\bm
S\cdot[\bm\pi\times\bm E]-\bm S\cdot[\bm E\times
\bm\pi]\bigr)\biggr\}  \\ \left.
-\rho_3\left\{\left(g-2+\frac{2m}{\epsilon'}\right), \bm S\cdot\bm
B\right\}\right.\\
\left.+\rho_3\frac{g-2}{4}\left\{\frac{1}{\epsilon'(\epsilon'+m)},
\bigl\{\bm S\cdot\bm\pi,(\bm\pi\cdot\bm B+\bm
B\cdot\bm\pi)\bigr\}\right\}\right]\\+
\frac{e\eta}{8m}\left[\left\{\frac{1}{\epsilon'}, \bigl(\bm
S\cdot[\bm B\!\times\!\bm\pi]-\bm S\cdot[\bm\pi\!\times\!\bm
B]\bigr)\right\}-4\rho_3\bm S\cdot\bm E \right. \\ \left.
+\frac{\rho_3}{2}\left\{\frac{1}{\epsilon'(\epsilon'+m)},
\bigl\{\bm S\cdot\bm\pi,(\bm\pi\cdot\bm E+\bm
E\cdot\bm\pi)\bigr\}\right\}\right],
\end{array} \label{eq20} \end{equation}
where $\epsilon'=\sqrt{m^2+\bm\pi^2}$. In this Hamiltonian, terms
neither bilinear in the field strengths nor those containing
derivatives of these strengths are taken into account. Commutators
and anticommutators are defined by $[\dots,\dots]$ and
$\{\dots,\dots\}$, respectively. Equation (\ref{eq20}) generalizes
the corresponding one derived in Ref. \cite{JETP} without
allowance for the EDM terms.

\section{Equations of motion and their classical limit}

Equation (\ref{eq20}) allows us to derive quantum-mechanical
equations of motion and then obtain their classical limit. Such
equations are defined by the commutators of the FW Hamiltonian
with appropriate operators:
\begin{equation}\begin{array}{c}
\frac{d\bm\pi}{dt}=\frac{i}{\hbar}[{\cal
H}_{FW},\bm\pi] + \frac{\partial\bm\pi}{\partial
t}=\frac{i}{\hbar}[{\cal H}_{FW},\bm\pi]-e\frac{\partial\bm
A}{\partial t}, \\
\frac{d\bm\Pi}{dt}=\frac{i}{\hbar}[{\cal
H}_{FW},\bm\Pi]=\frac12\left(\bm\Omega\times\bm\Pi-\bm\Pi\times\bm\Omega\right),
\label{eqme} \end{array} \end{equation} where $\bm\Omega$ is the
operator of angular velocity of spin motion.

The operator equation of spin motion has the form:
\begin{equation}
\begin{array}{c}
\frac{d\bm{\Pi}}{dt} =\frac{1}{4}\Biggl\{\biggl(\frac{\mu_0m}
{\epsilon '+m}+\mu'\biggr)\frac{1}{\epsilon
'},\biggl(\bm{\Pi}\times[\bm E\times\bm \pi]\\
-\bm{\Pi}\times[\bm \pi\times\bm E]\biggr)\Biggr\}+
\frac{1}{2}\biggl\{\left( \frac{\mu_0m}{\epsilon
'}+\mu'\right),[\bm\Sigma\times\bm B]\biggr\}\\
-\frac{\mu'}{4}\left\{\frac{1}{\epsilon '(\epsilon '+m)},\biggl(
[\bm\Sigma\times\bm \pi](\bm \pi\cdot\bm B)+(\bm B\cdot\bm \pi)
[\bm\Sigma\!\times\!\bm
\pi]\biggr)\right\}\\-\frac{d}{4}\left\{\frac{1}{\epsilon
'},\left(\bm{\Pi}\times[\bm B\times\bm \pi]-\bm{\Pi}\times[\bm
\pi\times\bm B]\right)\right\}+d[\bm\Sigma\times\bm
E]\\-\frac{d}{4}\left\{\frac{1}{\epsilon '(\epsilon '+m)},\biggl(
[\bm\Sigma\times\bm \pi](\bm \pi\cdot\bm E)+(\bm E\cdot\bm \pi)
[\bm\Sigma\!\times\!\bm \pi]\biggr)\right\},
\end{array} \label{eq14sp1} \end{equation}
where $\mu_0=e/m,~\bm\Sigma={\cal{I}}\bm S$ (${\cal{I}}$ is the
unit $2\times2$ matrix). In this equation, second-order terms in
spin are not taken into account. Their contribution into the
Hamiltonian is 
considered below. Equation (\ref{eq14sp1}) is fully consistent
with the corresponding one for spin-1/2 particles with the EDM
\cite{RPJ}. Finding its classical limit reduces to the replacement
of operators by respective classical quantities \cite{JINRLlimit}
and results in
\begin{equation} \begin{array}{c}
\frac{d\bm s}{dt}=\frac{e}{m}\Biggl\{\frac{1}{\epsilon
'}\left(\frac{g-2}{2}+\frac{m} {\epsilon '+m}\right)\left[\bm
s\times[\bm E\times
\bm\pi]\right]\\+\left(\frac{g-2}{2}+\frac{m}{\epsilon'}\right)[\bm
s\times\bm B] -\frac{g-2}{2\epsilon'(\epsilon'+m)} [\bm s\times\bm
\pi](\bm{\pi}\cdot\bm B)\\-\frac{\eta}{2\epsilon'} \left[\bm
s\times[\bm B\times \bm\pi]\right] +\frac{\eta}{2}[\bm s\times\bm
E]\\-\frac{\eta}{2\epsilon'(\epsilon'+m)} [\bm s\times\bm
\pi](\bm{\pi}\cdot\bm E)\Biggr\}. \end{array} \label{Ieqt}
\end{equation} Equation (\ref{Ieqt}) coincides with the
respective classical one \cite{Nelson:1959zz} which generalizes
the Thomas-Bargmann-Michel-Telegdi equation \cite{T,BMT}.

This result perfectly proves self-consistency of the relativistic
wave equations for spin-1 particles which was called in question
for a long time (see Refs. \cite{VSM,TY} and references therein).
Their self-consistency is also confirmed by the form of the
quantum-mechanical equation of particle motion:
\begin{equation}   \begin{array}{c}
\frac{d\bm \pi}{dt}=e\bm
E+\rho_3\frac{e}{4}\Biggl\{\frac{1}{\epsilon'},
\biggl([\bm\pi\times\bm B]-[\bm B\times\bm\pi]\biggr)\Biggr\} \\+
\frac 14\Biggl\{\biggl(\frac{\mu_0m}{\epsilon'+m}+\mu'\biggr)
\frac{1}{\epsilon'},\nabla\biggl(\bm\Sigma\cdot[\bm
E\!\times\!\bm\pi]- \bm\Sigma\cdot[\bm\pi\!\times\!\bm
E]\biggr)\Biggr\}
\\+
\frac{1}{2}\Biggl\{\biggl(\frac{\mu_0m}{\epsilon'}+\mu'\biggr),
\nabla(\bm\Pi\cdot\bm B)\Biggr\} \\-
\frac{\mu'}{8}\Biggl\{\frac{1}{\epsilon'(\epsilon'+m)},\biggl\{(\bm{\Pi}
\!\cdot\!\bm\pi),\nabla\Bigl(\bm\pi\!\cdot\!\bm{B}\!+\!\bm{B}
\!\cdot\!\bm\pi\Bigr)\Biggr\}\Biggr\}.
\end{array} \label{eqVM} \end{equation} In Eq. (\ref{eqVM}), the terms dependent
on the EDM are omitted. This equation agrees with the
corresponding one describing spin-1/2 particles \cite{JMP}, and it is also in accord with classical theory. 
Spin-dependent terms define a relativistic Stern-Gerlach force.
For spin-1 particles in a uniform magnetic field, the
quantum-mechanical equations of motion have been derived in Ref.
\cite{EPJC}.

\section{Scalar and tensor polarizabilities of pointlike particles}

The W boson being a charged pointlike (structureless) spin-1
particle can be described by the PCS equations. Such a particle
possesses some electric and magnetic moments. The W boson may have
an AMM (see Refs. \cite{VReSc,LarLeMa,Revista} and references
therein). This AMM is defined by radiative corrections. The
quantum mechanics allows us to derive other moments of the
pointlike particle with the definite $g$ factor. Of course,
moments of pointlike particles can be affected by radiative
corrections, and moments of pointlike and composed particles can
significantly differ.

The quadrupole and contact interactions of a charged structureless
spin-1 particle possessing the AMM were first determined by Young
and Bludman \cite{YB} in the nonrelativistic approximation. In
Ref. \cite{JETP}, a relativistic description of these interactions
has been made. Pomeransky and Khriplovich \cite{PK} have obtained
relativistic expressions for the quadrupole interaction of
arbitrary-spin particles (without analysis of the contact
interaction). The nonrelativistic formula for the quadrupole and
contact interactions of a charged structureless spin-1 particle is
\begin{equation} \begin{array}{c}
W= \frac{e(g-1)}{4m^2}(S_iS_j+S_jS_i) \frac{\partial E_i}{\partial
x_j}- \frac{e(g-1)}{2m^2}\nabla\cdot\bm E.
\end{array} \label{eq21} \end{equation}
The corresponding quadrupole moment is equal to $Q=-e(g-1)/m^2$
\cite{YB,JETP,PK}.

The nonrelativistic FW transformation of the initial Young-Bludman
Hamiltonian (\ref{eq20}) makes it possible to
determine the 
polarizabilities defined as follows:
\begin{equation} \begin{array}{c}
\Delta{\cal H}_{FW}= -\frac12\alpha_SE^2\!-\!\frac12\beta_SB^2
\!-\!\alpha_T(\bm S\!\cdot\!\bm E)^2\!-\!\beta_T(\bm S\!\cdot\!\bm
B)^2.
\end{array} \label{eqT} \end{equation}
Here $\alpha_S$ and $\beta_S$ are the scalar electric and magnetic
polarizabilities, and $\alpha_T$ and $\beta_T$ are the tensor
electric and magnetic ones. The related terms in the
nonrelativistic FW Hamiltonian calculated for the first time read
\begin{equation} \begin{array}{c}
\Delta{\cal H}_{FW}=\rho_3\frac{e^2\hbar^2(g\!-\!1)^2}{2m^3}\bm
E^2-\rho_3\frac{e^2\hbar^2(g\!-\!1)^2}{2m^3}(\bm S\cdot\bm E)^2\\
-\rho_3\frac{e^2\hbar^2}{8m^3}\left[(g-1)^2+3\right](\bm S\cdot\bm
B)^2.
\end{array} \label{eqemp} \end{equation}

For positive-energy states, the polarizabilities are given by
\begin{equation} \begin{array}{c}
\alpha_S=-\frac{e^2\hbar^2(g-1)^2}{m^3}, ~~~
\alpha_T=\frac{e^2\hbar^2(g-1)^2}{2m^3}, \\
\beta_T=\frac{e^2\hbar^2}{8m^3}\left[(g-1)^2+3\right],
\end{array} \label{eqplriz} \end{equation}
and the scalar magnetic polarizability is zero. The tensor
electric and magnetic polarizabilities of spin-1 particles without
the AMM are equal to each other. The nonzero AMM brings a
difference between them.

\section{Discussion and summary}

The present work shows that the PCS equations can be added by the
EDM-dependent term. Further transformations allow us to obtain the
self-consistent Hamiltonians in the Sakata-Taketani and FW
representations. The classical limit of derived quantum-mechanical
equations of motion coincides with corresponding classical ones.
These results demonstrate self-consistency of quantum mechanics of
spin-1 particles and create the sufficient quantum-mechanical
basis for the planned deuteron EDM experiment
\cite{Semertzidis:2003iq,dEDM,Lehrach:2012eg}.

The deuteron EDM experiment in storage rings is very important.
This experiment is not only complementary to other EDM searches,
but for some potential sources of EDMs, it is superior (see Ref.
\cite{dEDM} and references therein).

The calculation of the scalar and tensor electric and magnetic
polarizabilities of pointlike spin-1 particles has been fulfilled
for the first time. The scalar magnetic polarizability happens to
be zero. The tensor electric and magnetic polarizabilities of a
spin-1 particle with the zero AMM are equal to each other.
Equation (\ref{eqplriz}) defines the parameters of the W boson
being a charged structureless spin-1 particle. In particular,
$\alpha_S=-1.1\times10^{-10}~{\rm
fm}^3,~\alpha_T=\beta_T=5.4\times10^{-11}~{\rm fm}^3$ for $g=2$
and $\alpha_S=-9.9\times10^{-11}~{\rm
fm}^3,~\alpha_T=5.0\times10^{-11}~{\rm
fm}^3,~\beta_T=5.3\times10^{-11}~{\rm fm}^3$ for \cite{Revista}
$g-2=-4.05\times10^{-2}$. The polarizabilities of composed spin-1
particles are much greater. For example, the tensor electric and
magnetic polarizabilities of the deuteron are of the order of
$10^{-2}\div10^{-1}~{\rm fm}^3$ \cite{CGSJLFP}, while the
corresponding values for a pointlike particle of the same mass
calculated by using Eq. (\ref{eqplriz}) are of the order of
$10^{-6}~{\rm fm}^3$. We can note an importance of allowance for
the tensor polarizabilities in the EDM experiment (see Ref.
\cite{dEDMtensor} and references therein).

\section*{Acknowledgements}

The work was supported by the Belarusian Republican Foundation for
Fundamental Research (Grant No. $\Phi$12D-002).


\end{document}